\documentclass[11pt]{iopart}
\usepackage{latexsym,epsfig}
\usepackage{bm}
\usepackage{amssymb,amsthm}

\newtheorem*{Lem}{Lemma}

\begin{document}

\title{Cotton flow}

\author{Ali Ula\c{s} {\"O}zg{\"u}r Ki\c{s}isel\dag, {\"O}zg{\"u}r Sar{\i}o\u{g}lu\ddag 
and Bayram Tekin\ddag}

\address{\dag Department of Mathematics, Faculty of Arts and  Sciences,\\
        Middle East Technical University, 06531, Ankara, Turkey}

\address{\ddag Department of Physics, Faculty of Arts and  Sciences,\\
        Middle East Technical University, 06531, Ankara, Turkey}

\ead{akisisel@metu.edu.tr, sarioglu@metu.edu.tr and btekin@metu.edu.tr}

\date{\today}

\begin{abstract}
Using the conformally invariant Cotton tensor, we define a geometric flow,
the {\it Cotton flow}, which is exclusive to three dimensions. This flow tends
to evolve the initial metrics into conformally flat ones, and is
somewhat orthogonal to the Yamabe flow, the latter being a flow within a
conformal class. We define an entropy functional, and study the flow of nine 
homogeneous spaces both numerically and analytically. In particular, we show that 
the arbitrarily deformed homogeneous 3-sphere flows into the round 3-sphere. Two 
of the nine homogeneous geometries, which are degenerated by the Ricci flow, are 
left intact by the Cotton flow. 
\end{abstract}



\maketitle

\section{Introduction}
In this paper, inspired by Hamilton's work on the Ricci flow \cite{hamilton1}
that culminated in Perelman's proof of Thurston's geometrization conjecture and the 
longstanding Poincar\'{e} conjecture \cite{perelman1, perelman2}, we introduce a new flow,
based on the conformally invariant Cotton tensor. This flow is exclusive to three
dimensions and its fixed points are the conformally flat metrics.

Historically, the Ricci flow equations seem to have first arisen in a work by 
Friedan \cite{frie} that deals with the renormalization group (RG)
flow in a 2-dimensional nonlinear sigma model. Just like any other coupling
in quantum field theories, the Riemannian metric, considered
as a coupling `constant' between the sigma model fields defined on a
manifold $M$, runs as the energy scale changes:
\begin{equation}\label{Ricciflow}
\beta_g = \partial_t g_{ij} = k \alpha^{\prime} R_{ij},
\end{equation}
where $k$ is a constant uniquely fixed by the 1-loop result, and
$\alpha^{\prime}$ is a dimensionful parameter (such as the tension). Of
course, in general, the beta function for the RG flow of the metric will
have infinitely many terms, including the Ricci tensor. In fact, in
\cite{frie}, the 2-loop result with a Riemann-square correction is
given. In string theory, which is again a two dimensional sigma model
with additional fields, gravitation (including General Relativity as
the lowest order approximation) that governs the dynamics of the target space, is 
derived by the requirement
that the beta function vanishes as a result of Weyl invariance on the
worldsheet. Since a non-anomalous Weyl symmetry is crucial for string theory,
in the physics literature, it is common practice to set $\beta_g=0$ and consequently, 
end up not with a flow, but with a usual gravity theory (containing
perhaps some additional fields and higher curvature terms). On the other
hand, Hamilton \cite{hamilton1} {\it introduced} the Ricci flow as a means to 
deform a given metric on a manifold and developed a programme aiming to prove the Poincar\'{e} conjecture. In particular, he classified closed 3-manifolds of positive Ricci 
curvature.

Obviously, one could introduce many different flows, or depending on one's
point of view, derive flows that come from the beta functions of two dimensional
non-linear sigma models. As examples of the former, see \cite{eels,hamilton2,cqgconf}.
For an example of the latter, see \cite{gegenberg} where a string inspired flow with scalar, Maxwell and anti-symmetric tensor fields is studied.

Our motivation was to find a flow whose critical
points are the conformally flat spaces, and this is achieved in three
dimensions with a unique 2-tensor called the Cotton tensor. We study the behaviour of the nine homogeneous geometries under this flow in detail. Four of these are fixed points of our flow. [We remark that the eight geometries appearing in Thurston's geometrization conjecture are maximally symmetric; viewed in this way, two of the nine ($Isom(\mathbb{R}^{2})$ and $\mathbb{R}^{3}$) collapse to the same maximally symmetric geometry.] 

Interesting physics could come out of the Ricci flow or other flows such as the one we introduce here, especially in Euclidean quantum gravity defined by the path integral. This is somewhat an unexplored territory. Yet, we can point out several works in this
direction: In \cite{headrick}, the Ricci flow is applied to four dimensional
black hole physics as a means to understand the phase
structure of this theory in Euclidean space. The Ricci flow
is also used \cite{samuel} to find inequalities in General Relativity regarding the
evolution of the area of a surface and the enclosed Hawking mass.
See \cite{woolgar} for a compilation of possible applications of the Ricci
flow in physics.

The outline of the paper is as follows: In section 2, we introduce the
Cotton flow and show how some geometric quantities evolve accordingly. In
section 3, we give an entropy functional which is non-decreasing under the
Cotton flow. This also gives us a gradient formulation of the Cotton flow. In
section 4, we show that the Cotton flow is in a certain sense orthogonal to the Yamabe flow. 
Section 5 is the bulk of our paper where we study the evolution of homogeneous spaces under the
Cotton flow both numerically and analytically. In the conclusion, we summarize our results 
and mention some topics for further study. Details of the computations regarding the homogeneous 
geometries are given in the appendices. 

\section{The Cotton flow}

In dimensions greater than or equal to 4, the Weyl tensor
determines whether or not a given manifold is locally conformally
flat. In 3-dimensions, the Weyl tensor vanishes identically, and the role of the Weyl tensor 
is played by the Cotton tensor \cite{Cot,Eis,adm,york,garcia}
\begin{equation}
C^{ij}=\frac{\epsilon^{imn}}{\sqrt{g}}\nabla_{m}(R^{j}_{\,n}-\frac{1}{4}\delta^{j}_{n}\mathcal{R}),
\end{equation}
where $\epsilon$ is a tensor density, in an orthonormal frame
$\epsilon^{123}=1$, $R_{ij}$ is the Ricci tensor, and
$\mathcal{R}$ the curvature scalar. [Note that the Cotton tensor is also
referred to as the `Cotton-York tensor' since York reintroduced \cite{york}
and used it extensively in the initial value formulation of General Relativity.]
Here and throughout, the signature of the space will be $(+,+,+)$, the indices will range
from $\{1,2,3\}$, and the sign convention for the Riemann tensor will follow from
$[\nabla_{i},\nabla_{j}]v_{k}=R_{ijk}\,^{l}v_{l}$. The tensor
$C^{ij}$ is covariantly conserved, symmetric and traceless (see
\cite{DJT}, where the Cotton tensor was used to introduce the
``topologically massive gravity''). The symmetry can be directly
seen from the following representation of $C^{ij}$:
\begin{equation}
C^{ij} = \frac{1}{2\sqrt{g}} \Big(
\epsilon^{imn} \nabla_{m} R^{j}_{\,n} + \epsilon^{jmn} \nabla_{m} R^{i}_{\,n} \Big) .
\end{equation}
The tensor $C^{i}\,_{j}$ is conformally invariant with weight $0$.

Let $M$ be a smooth $3$-manifold with a positive definite metric
$g_{ij}$.  We will consider the flow defined by the equation
\begin{equation}\label{Cottonflow}
\partial_{t}g_{ij}=\kappa C_{ij},
\end{equation}
where $\kappa$ is a positive constant (note that $t$ is not one of the coordinates of the manifold). 
Without loss of generality, we may scale $t$ to set $\kappa=1$. The reason for
taking $\kappa>0$ is to make homogeneous metrics on the 3-sphere converge to the round 
sphere rather than diverge from it. Since the right hand side is traceless, $\sqrt{g}$ is 
independent of $t$ although it could depend on the coordinates. In contrast, the Ricci tensor 
is not traceless in general, so a normalization factor is necessary for the Ricci flow (\ref{Ricciflow})
in order for the volume density or volume to be preserved.

Assuming that $g_{ij}$ flows under the evolution equation (\ref{Cottonflow}), one can compute how some other geometric quantities evolve:
\begin{eqnarray}\label{quantities}
\partial_{t}\Gamma^{i}\,_{jk}=\frac{1}{2}g^{il}\big(\nabla_{j}C_{kl}+\nabla_{k}C_{jl}-\nabla_{l}C_{jk}\big), \nonumber\\
\partial_{t}R_{ij}=3 R_{l(i}C_{j)}\,^{l} - R^{lm}C_{lm}g_{ij} - \frac{1}{2}\mathcal{R}C_{ij}-\frac{1}{2}\nabla^{2}C_{ij},\nonumber\\
\partial_{t}\mathcal{R}=-C^{ij}R_{ij},\nonumber\\
\partial_{t} C^{ij}=\epsilon^{mk(i}\nabla_{|m|}\partial_{t} R^{j)}\,_{k}+ \epsilon^{mk(i}\partial_{t} \Gamma^{j)}\,_{mq}R^{q}\,_{k},
\nonumber\\
\partial_{t}(R_{jk}R^{jk})=4R_{jl}C^{l}\,_{k}R^{jk}-3\mathcal{R}R_{jk}C^{jk}-R^{jk}\nabla^{2}C_{jk},
\end{eqnarray}
where the usual shorthand notation for symmetrization is used: \( A_{(i|j|k)} = (A_{ijk}+A_{kji})/2 \).

The Cotton flow equations form a system of third order, nonlinear partial differential equations. About a flat background, the linearization of the system, setting  $g_{ij}=\delta_{ij}+h_{ij}$, gives
\begin{equation}
\partial_{t}h^{ij}=-\frac{1}{4}\partial^{2}\partial_{k}(\epsilon^{ikl}h^{j}_{l}+\epsilon^{jkl}h^{i}_{l})+\partial^{i}\xi^{j}+\partial^{j}\xi^{i},
\end{equation}
where
\begin{equation}
\xi^{i}=-\frac{1}{4}\epsilon^{ikl}\partial_{k}\partial^{n}h_{nl}.
\end{equation}
and $\partial^{2}$ denotes the Laplacian in flat space. The $\xi$ terms can be viewed as coming from a diffeomorphism of the manifold. Unlike the Ricci flow case, whose linearized form is of heat equation type in the highest order after DeTurck's modification by a diffeomorphism \cite{DT}, our equation is of third order. In this respect, it seems impossible to apply results from elliptic operator theory directly for the short time existence problem. Understanding the conditions for the initial metric under which the equations are locally well-posed seems to be a delicate problem, which will not be studied in this paper.

\section{Cotton entropy}\label{entropy}

In order to show that the Ricci flow can be regarded as a gradient flow, Perelman \cite{perelman1} defined an {\it entropy functional} $\mathcal{F}(g,f)=\int_{M}(R+|\nabla f|^{2})e^{-f}
\sqrt{g}d^{3}x$, where $g$ denotes the metric, and $f$ is a scalar field. He proved that if $\partial_{t}f=\partial_{t}\ln{\sqrt{g}}$ and $g$ obeys the Ricci flow equations, then $\mathcal{F}$ is a (non-strictly) increasing functional. For the Cotton flow, an analogous functional is the well known gravitational Chern-Simons action \cite{DJT}:
\begin{equation}
\mathcal{F}(g)=-\frac{1}{2}\int_{M} \epsilon^{ijk}\,\Gamma^{l}\,_{im}\Big(\partial_{j}\Gamma^{m}\,_{kl}+\frac{2}{3}\Gamma^{m}\,_{jn}\Gamma^{n}\,_{kl}\Big) d^{3}x.
\end{equation}
Note the absence of an extra scalar field, in contrast to the case of the Ricci flow. The functional $\mathcal{F}$ is conformally invariant, and up to a boundary term, diffeomorphism invariant. Defining $\partial_{t}g_{ij}=v_{ij}$, after a straightforward but lengthy computation, one obtains
\begin{equation}
\frac{d\mathcal{F}}{dt}=\int_{M}\sqrt{g}v_{ij}C^{ij}d^{3}x.
\end{equation}
The choice $v_{ij}=C_{ij}$ gives the steepest descent, and leads to the Cotton flow $\partial_{t}g_{ij}=C_{ij}$. The functional $\mathcal{F}(g)$ is increasing if $g$ evolves under the Cotton flow. The Cotton entropy $\mathcal{F}(g)$ is constant if and only if $C_{ij}$ is identically zero, which means that $M$ is locally conformally flat at all points.

\section{Cotton flow and the Yamabe flow}

One can define various flows in the space of metrics. One of these
is the Yamabe flow defined by the equation \cite{hamilton2}
\begin{equation}
\partial_{\tau}g_{ij}=-\mathcal{R}g_{ij}.
\end{equation}
Under this flow, the conformal class of a metric does not change.
The Yamabe flow was introduced in order to solve Yamabe's
conjecture \cite{hamilton2}, which states that any metric is conformally equivalent to a metric
with constant scalar curvature. The Cotton flow has a somewhat
complementary behaviour since by definition it changes the
conformal class unless the metric is conformally flat. We will now
make this behaviour precise in a sense, by computing the
commutator of the vector fields for these two flows.

Let $t$ and $\tau$ denote, respectively, the parameters for the
Cotton and the Yamabe flows. Then
\begin{equation}
[\partial_{t}, \partial_{\tau}] g_{ij} = \partial_{t} \partial_{\tau} g_{ij} 
- \partial_{\tau} \partial_{t} g_{ij}
= - \partial_{t} (\mathcal{R}g_{ij}) - \partial_{\tau} C_{ij}.
\end{equation}
Since $C^{i}\,_{j}$ is conformally invariant,
$\partial_{\tau}C^{i}\,_{j}=0$. From here, it follows that
$\partial_{\tau}C_{ij}=-\mathcal{R}C_{ij}$. Using the third
equality of (\ref{quantities}), one gets
$\partial_{t}(\mathcal{R}g_{ij})=\mathcal{R}C_{ij}-R^{kl}C_{kl}g_{ij}$.
Therefore,
\begin{equation}
[\partial_{t},\partial_{\tau}]g_{ij}=R^{kl}C_{kl}g_{ij}.
\end{equation}
Since the result is proportional to the metric, the commutator of
the two flows gives another flow preserving the conformal class,
just like the Yamabe flow.

\section{Cotton flow on homogeneous 3-manifolds}\label{hofl}

In this section, we will be interested in studying the behaviour
of the Cotton flow on homogeneous three manifolds. There are nine 
such homogeneous geometries: $SU(2)$,
$SL(2,\mathbb{R})$, $Isom(\mathbb{R}^{2})$, $Solv$, $Nil$,
$\mathbb{R}^{3}$, $H^{3}$, $S^{2}\times \mathbb{R}$, $H^{2}\times
\mathbb{R}$. As explained in the introduction, they give rise to the eight maximally symmetric geometries which are the basic building blocks
that take the stage in Thurston's geometrization conjecture for
three manifolds \cite{Thu}. This conjecture, which includes the
Poincar\'{e} conjecture as a special case, was settled by Perelman's
work \cite{perelman1,perelman2}. 

The first six of these nine geometries can be obtained by the following construction: Suppose that $G$ is a $3$-dimensional, unimodular, simply
connected Lie group, and that $g$ is a left-invariant metric on
$G$. Let $M$ be a compact 3-manifold obtained as the quotient of
$G$ with respect to a discrete subgroup. The analysis of the Ricci
flow on these manifolds was carried out in detail in \cite{IJ,KM}
(see also \cite{CK}). [A similar analysis for cross curvature flow was carried out in
\cite{caoni}.] As in the case of the Ricci flow, the Cotton
flow equations on a homogeneous 3-manifold reduce to a set of three
coupled ordinary, nonlinear, autonomous differential equations.

We now proceed to find this system of differential equations. In \cite{Mil}, it is shown that there exists an orthogonal left-invariant frame
$\{F_{1},F_{2},F_{3}\}$ on $G$ such that $[F_{1},F_{2}]=2\nu F_{3}$, $[F_{2},F_{3}]=2\lambda F_{1}$,
$[F_{3},F_{1}]=2\mu F_{2}$ where $\mu,\nu,\lambda\in\{-1,0,1\}$ (see also \cite{CK}).
Let $\{\omega^{1},\omega^{2},\omega^{3}\}$ be the dual basis of 1-forms to $\{F_{1},F_{2},F_{3}\}$.
In this basis, the metric $g$ takes the form
\begin{equation}\label{homogenousmetric}
g=A(t)\omega^{1}\otimes\omega^{1}+B(t)\omega^{2}\otimes\omega^{2}+C(t)\omega^{3}\otimes\omega^{3}.
\end{equation}
Defining $e^{1}=\sqrt{A}\omega^{1}, e^{2}=\sqrt{B}\omega^{2},e^{3}=\sqrt{C}\omega^{3}$ we obtain an orthonormal coframe,
with respect to which the metric takes the form $g=e^{1}\otimes e^{1}+e^{2}\otimes e^{2}+e^{3}\otimes e^{3}$. One has
\begin{equation}
d\omega^{1}=2\lambda \omega^{2}\wedge\omega^{3}, \quad
d\omega^{2}=2\mu \omega^{3}\wedge\omega^{1}, \quad
d\omega^{3}=2\nu \omega^{1}\wedge\omega^{2}.
\end{equation}
The Cotton flow equations can be written as $\partial_{t}e^{a}=*C^{a}$, where $C^{a}$ is the Cotton $2$-form and $*$
denotes the Hodge dual. Since the Cotton flow preserves the volume density, from now on we assume that $ABC=1$. Referring the reader to the appendix for the details of the computation, we state the equations which take the form
\begin{eqnarray}\label{homogenousflow}
\frac{dA}{dt}&=4A[-\lambda^{2}A^{2}(\mu B+\nu C-2\lambda A)-(\mu B+\nu C)(\mu B-\nu C)^{2}],\nonumber\\
\frac{dB}{dt}&=4B[-\mu^{2}B^{2}(\nu C+\lambda A-2\mu B)-(\nu C+\lambda A)(\nu C-\lambda A)^{2}],\nonumber\\
\frac{dC}{dt}&=4C[-\nu^{2}C^{2}(\lambda A+\mu B-2\nu C)-(\lambda
A+\mu B)(\lambda A-\mu B)^{2}].
\end{eqnarray}

The entropy functional $\mathcal{F}$ defined in section \ref{entropy} can be alternatively computed using
\[\mathcal{F}=-\frac{1}{4}\int \Big(\omega^{a}\,_{b}\wedge d\omega^{b}\,_{a}+\frac{2}{3}\omega^{a}\,_{b}\wedge \omega^{b}\,_{c}\wedge \omega^{c}\,_{a}\Big),\]
which for our case yields
\begin{equation}
\mathcal{F}=\int [4\lambda\mu\nu
-2(\lambda A+\mu B-\nu C)(\lambda A-\mu B+\nu C)(-\lambda A+\mu B+\nu C)] *1,
\end{equation}
where $*1= e^{1} \wedge e^{2} \wedge e^{3}$.  
Recall that by the result in section \ref{entropy}, $\mathcal{F}$ is a non-decreasing functional. [The name ``entropy'' may be misleading since $\mathcal{F}$ is not necessarily nonnegative. In fact, the maximum value of $\mathcal{F}$ in the $SU(2)$ case below is $-2$. All that matters is that $\mathcal{F}$ is non-decreasing.] The following simpler functional, which was obtained by manipulating (\ref{homogenousflow}), also acts like an entropy in {\it certain} cases, and will also be useful in what follows:
\begin{equation}
f(t)=\frac{1}{8}\Big(\frac{\mu^{2}\nu^{2}}{A^{2}}+\frac{\lambda^{2}\nu^{2}}{B^{2}}+\frac{\lambda^{2}\mu^{2}}{C^{2}}\Big).
\end{equation}
Note also that, using (\ref{homogenousflow})
\begin{eqnarray}
\frac{df}{dt}=&\frac{\mu^{2}\nu^{2}}{A^{2}}(\mu B+\nu C)(\mu B-\nu C)^{2}+\frac{\lambda^{2}\nu^{2}}{B^{2}}(\nu C+\lambda A)(\nu C-\lambda A)^{2}\nonumber\\
&+\frac{\lambda^{2}\mu^{2}}{C^{2}}(\lambda A+\mu B)(\lambda A-\mu B)^{2}.
\end{eqnarray}

\subsection{SU(2)}
When we take $\lambda=\mu=\nu=-1$, the Lie group $G$ is isomorphic
to $SU(2)$, which is homeomorphic to $S^{3}$ as a topological
manifold. The proof that the flow equations (\ref{homogenousflow})
take any initial homogeneous metric to the round metric $A=B=C=1$
on $S^{3}$ is remarkably simple. Since $\lambda=\mu=\nu=-1$, one has
$f(t)\geq 3/8$, $df/dt\leq 0$, and in both cases equality
occurs if and only if $A=B=C=1$. This proves the claim.

It seems impossible to find an exact analytical solution to this system. Therefore
 we carry out an estimate analysis similar to that in \cite{IJ}. From (\ref{homogenousflow}), one gets
\begin{equation}
\frac{d(A-B)}{dt}=8(B-A)[2(B^{3}+B^{2}A+BA^{2}+A^{3})-C(A+B)^{2}-C^{3}].
\end{equation}
Since the right hand side vanishes when $A=B$, by the uniqueness of
solutions, the ordering between $A$ and $B$, and likewise the
ordering of the three variables cannot change during the flow.
Therefore, without loss of generality, we may assume that $A\geq
B\geq C$ due to the symmetry of the equations. From the last equation of (\ref{homogenousflow}),
it is clear that $C$ is not decreasing, so $A\geq B\geq C\geq
C_{0}$ where $C_{0}$ is the initial value of $C$. Moreover,
\begin{eqnarray}
\frac{d(A-C)}{dt}&=-8(A-C)[2C^{3}+2C^{2}A+2CA^{2}+2A^{3}-B(A+C)^{2}-B^{3}]\nonumber\\
&\leq -8(A-C)[2C^{3}+2C^{2}A+2CA^{2}+2A^{3}-A(A+C)^{2}-A^{3}]\nonumber\\
&\leq -24C_{0}^{3}(A-C).
\end{eqnarray}
We deduce that $A-C\leq (A_{0}-C_{0})\exp(-24C_{0}^{3}t)$. Since
the difference between the greatest and smallest of $A,B,C$ decays
at least at this rate to $0$, $(A,B,C)$ also converges at least at
this rate to $(1,1,1)$.
\begin{figure}[ht]
\begin{center}
\epsfig{file=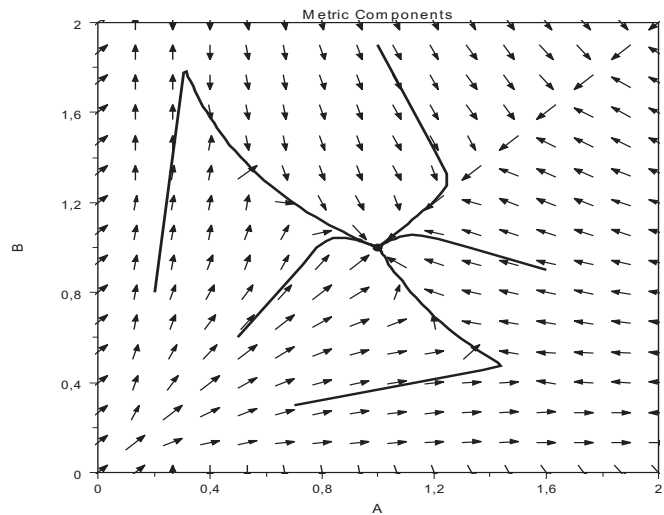,height=6.9cm,width=8.9cm} \caption{The
solid lines indicate flows, for various initial metrics on $S^{3}$: $(A_{0},B_{0})=$ $(0.2,0.8),$ $(0.5,0.6),$ $(0.7,0.3),$ $(1,1.9),$ $(1.6,0.9)$.
The horizontal axis is for $A(t)$ and the vertical axis is for
$B(t)$. $C(t)$ is not depicted as it runs according to $C=1/AB$.}
\end{center}
\end{figure}

Linearizing the system around the fixed point $(A,B)=(1,1)$ after setting $C=1/AB$ gives
\begin{equation}
\frac{dA}{dt}=-24A, \quad \frac{dB}{dt}=-24B,
\end{equation}
which is in accordance with the estimate above since $\max(C_{0})=1$.

One can alternatively solve the equations numerically. Starting from various initial values for $(A_{0},B_{0},C_{0})$, a sketch of the solution curves can be seen in figure 1. 

\subsection{$SL(2,\mathbb{R})$}

When we take $\lambda=\mu=-1$, $\nu=1$, the group $G$ is
isomorphic to the universal cover of $SL(2,\mathbb{R})$. From (\ref{homogenousflow}) one
gets
\begin{equation}
\frac{d(A-B)}{dt}=4(B-A)[2(B^{3}+B^{2}A+BA^{2}+A^{3})+C(B^{2}+A^{2})+2+C^{3}].
\end{equation}
Since the right hand side is $0$ when $A=B$, we deduce that the
ordering between $A$ and $B$ does not change throughout the flow.
Without loss of generality, assume $A\geq B$ (note that the
equations for $A$ and $B$ are symmetric). The equation for $A$ reads 
\begin{eqnarray}\label{ineqA}
\frac{dA}{dt}&=4A[A^{2}B-A^{2}C-2A^{3}+B^{3}+B^{2}C-BC^{2}-C^{3}]\nonumber\\
&\leq -4(ABC^{2}+AC^{3})=-4(C+AC^{3}).
\end{eqnarray}
This shows that $A$ is strictly decreasing (However $B$ is not necessarily decreasing). The equation for $C$ gives a lower bound 
\begin{equation}
\frac{dC}{dt}=4C[C^{2}(A+B+2C)+(A+B)(A-B)^{2}]\geq 8C^{4}. 
\end{equation}
\begin{figure}[ht]
\begin{center}
\epsfig{file=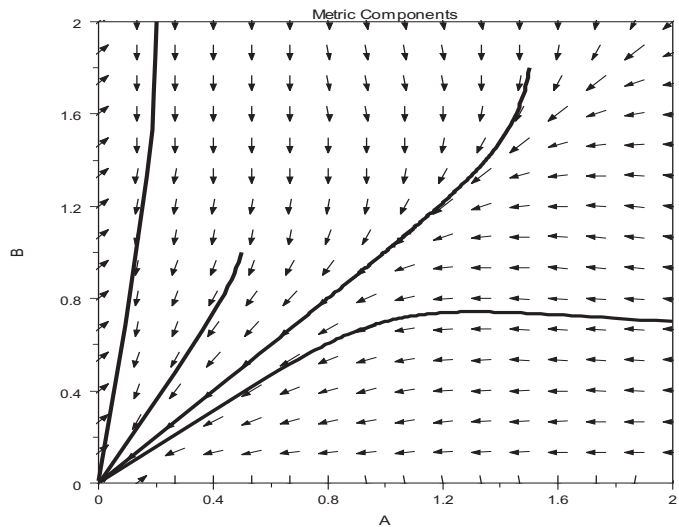, height=6.9cm,width=8.9cm} \caption{The
solid lines indicate flows, for various initial metrics on a
homogeneous geometry modelled on $SL(2,\mathbb{R})$:
$(A_{0},B_{0})=$ $(0.5,1),$ $(2,0.7),$ $(1.5,1.8),$ $(0.2,2)$.}
\end{center}
\end{figure}
Therefore, $C\to \infty$ in a finite time $t_{f}$. We want to show that $A(t_{f})=B(t_{f})=0$. Since there exists a point $t_{1}$ of time after which $C\geq A+B$, we have $dC/dt\leq 16C^{4}$. For $t\geq t_{1}$ 
\begin{equation}
\frac{d(A^{5}C)}{dt}\leq -4A^{5}C^{4}\leq 0,
\end{equation}
which implies that $A(t_{f})=0$. Since $B(t_{f})\leq A(t_{f})$, we also get $B(t_{f})=0$.   
In the terminology of
\cite{IJ}, the geometry always tends to a ``cigar degeneracy'' in finite time. Some numerical solutions of the system are given in figure 2. 

\subsection{$Isom(\mathbb{R}^{2})$} \label{Isom}

When we take $\lambda=\mu=-1$, $\nu=0$, $G$ is isomorphic to the universal cover of the group of Euclidean motions of $\mathbb{R}^{2}$. From (\ref{homogenousflow}) one
obtains
\begin{equation}\label{dogru}
\frac{d(A-B)}{dt}=8(B^{4}-A^4), \quad
\frac{dC}{dt}=4C(A^{2}-B^{2})(A-B).
\end{equation}
From these two equations, one gets
\begin{equation}\label{isomdiff}
C(A-B)\frac{d(A-B)}{dt}+2\Big((B-A)^{2}+\frac{2}{C}\Big)\frac{dC}{dt}=0.
\end{equation}
The function $C^{3}$ is an integrating factor for this equation.
So, there is a second conserved quantity besides $ABC=1$:
\begin{equation}\label{conserved}
C^{4}(A-B)^{2} + \frac{8}{3}C^{3} = k,
\end{equation}
where $k$ is a constant.
Since there are two conserved quantities and three variables, the
phase space foliates into 1-dimensional leaves. From (\ref{dogru}), 
we see that the ordering between the symmetric variables $A$ and $B$ 
doesn't change during the flow. Assume without loss
of generality that $A\geq B$. Then,
\begin{eqnarray}
\frac{dA}{dt}=4A(A^{2}B-2A^{3}+B^{3})\leq 0, \\
\frac{dB}{dt}=4B(A^{3}+AB^{2}-2B^{3})\geq 0.
\end{eqnarray}
Thus $A$ is decreasing, $B$ is increasing, and they are both bounded away
from $0$. This implies that the flow exists for all $t$. If $B_{0}$ is the
initial value of $B$, then since $B\geq B_{0}$, from (\ref{dogru}) one gets
\begin{eqnarray}
\frac{d(A-B)}{dt}&=-8(A-B)(B^{3}+B^{2}A+BA^{2}+A^{3})\\
&\leq -32B_{0}^{3}(A-B) .
\end{eqnarray}
So, $A-B$ decays to $0$. In fact, using the conserved quantities, the equations can be integrated.
From (\ref{conserved}), it follows that \( (A-B)^{2}= \frac{k}{C^{4}}-\frac{8}{3C} \). Using
$AB=1/C$, one obtains \( (A+B)^{2}= \frac{k}{C^{4}}+\frac{4}{3C} \). Then, the second equation
in (\ref{dogru}) yields
\[ \frac{dC}{dt} = 4 (\frac{k}{C^{3}}-\frac{8}{3}) \sqrt{\frac{k}{C^{4}}+\frac{4}{3C}} \,. \]
Defining 
\[ C = \Big( \frac{3}{4} (u^{2}-k) \Big)^{1/3}, \]
one finds
\[ \frac{u^{2}-k}{3k-2u^{2}} \frac{du}{dt} = \frac{32}{3}, \]
whose integration leads to an implicit definition of $u(t)$ as
\[ - \frac{u}{2} + \frac{\sqrt{k}}{2\sqrt{6}} \tanh^{-1}(u \sqrt{2/3k}) = \frac{32}{3}t + l \]
and $l$ is another constant. The remaining metric functions follow easily as  
\begin{eqnarray*}
A = \frac{1}{2} \Big( \frac{3}{4} (u^{2}-k) \Big)^{-2/3} (u + \sqrt{3k-2u^{2}}) \,, \\
B = \frac{1}{2} \Big( \frac{3}{4} (u^{2}-k) \Big)^{-2/3} (u - \sqrt{3k-2u^{2}}) \,. 
\end{eqnarray*}
Since $A-B \to 0$, by (\ref{conserved}), the limiting value of $C$ as $t\to \infty$
is $(3k/8)^{1/3}$. In the limit, $A=B$ and the geometry becomes flat. Even 
though we have analytically solved the equations, we also present some numerical 
solutions of the system for various initial conditions in figure 3. 
\begin{figure}[ht]
\begin{center}
\epsfig{file=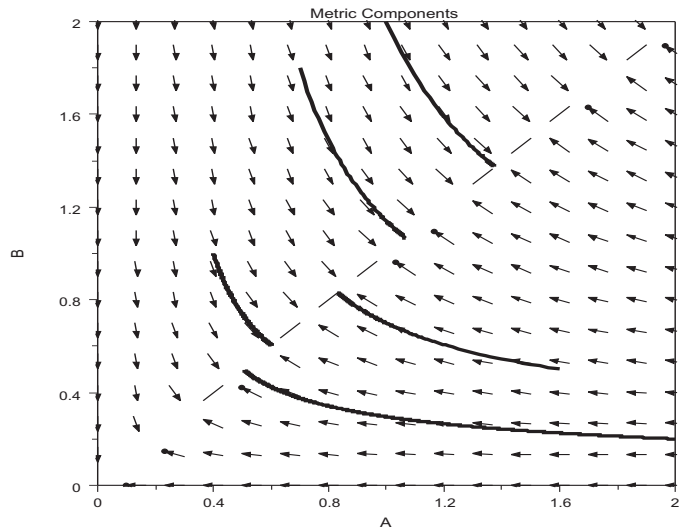, height=6.9cm,width=8.9cm} \caption{The
solid lines indicate flows, for various initial metrics on a
homogenous geometry modelled on $Isom(\mathbb{R}^{2})$:
$(A_{0},B_{0})=$ $(2,0.2),$  $(1,2),$ $(1.6,0.5),$ $(0.7,1.8),$ $(0.4,1)$.}
\end{center}
\end{figure}

The conserved quantity (\ref{conserved}) can be, more invariantly,
written in terms of the Ricci tensor and the scalar curvature as
\begin{equation}
-\frac{16^{4}}{6}\frac{\mathcal{R}^{3}Ric^{2}}{(3\mathcal{R}^{2}-Ric^{2})^{4}}=\mbox{const}.
\end{equation}
To get a unitless quantity, one should divide this by
$\mathrm{Vol}(M)^{2}$, if the manifold is of finite volume.

\subsection{$Solv$ geometry}

When we take $\lambda=-1$, $\mu=1$, $\nu=0$, the group $G$ is isomorphic to the group of 
isometries of the Minkowski plane. The equations (\ref{homogenousflow})
take the form:
\begin{eqnarray}
\frac{dA}{dt}=-4A(A^{2}(B+2A)+B^{3}), \quad
\frac{dB}{dt}=4B(B^{2}(A+2B)+A^{3}), \nonumber \\
\frac{dC}{dt}=-4C(B^{2}-A^{2})(A+B).
\end{eqnarray}
\begin{figure}[ht]
\begin{center}
\epsfig{file=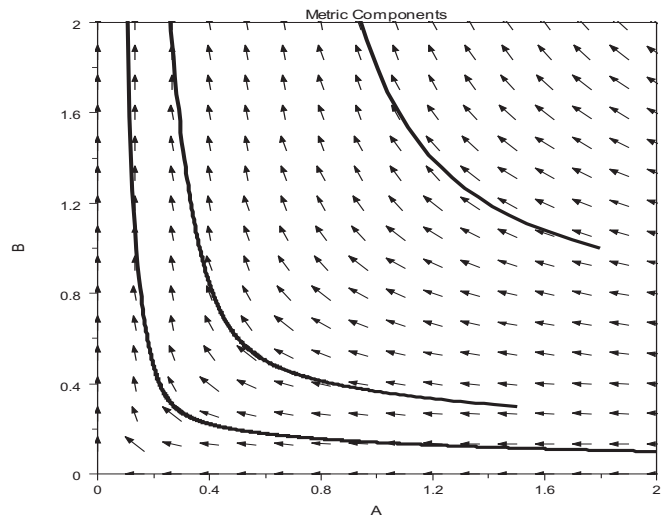, height=6.9cm,width=8.9cm} \caption{The
solid lines indicate flows, for various initial metrics on a {\it
Solv} geometry: $(A_{0},B_{0})=$ $(2,0.1),$ $(1.5,0.3),$ $(1.8,1)$.}
\end{center}
\end{figure}
The first two equations imply that $A$ is decreasing and $B$ is
increasing. In fact, $dB/dt \geq 8B^{4}$, which shows that $B \to \infty$ 
in finite time. As in section \ref{Isom}, one can find
a second conserved quantity which now reads
\begin{equation}
(A+B)^{2}C^{4} - \frac{8}{3}C^{3} = k, 
\end{equation}
where $k$ is a constant. Then, since $A$ is decreasing, $C$ must decay to $0$ at a rate
like $1/\sqrt{B}$. Recalling $ABC=1$, we see that $A$ also decays
to $0$ like $1/\sqrt{B}$. All geometries in this class thus tend
to a cigar degeneracy. Again, the equations can be completely integrated to give 
\begin{eqnarray*}
A = \frac{1}{2} \Big( \frac{3}{4} (k-u^{2}) \Big)^{-2/3} (\sqrt{3k-2u^{2}} - u) \,, \\
B = \frac{1}{2} \Big( \frac{3}{4} (k-u^{2}) \Big)^{-2/3} (\sqrt{3k-2u^{2}} + u) \,, \quad
C = \Big( \frac{3}{4} (k-u^{2}) \Big)^{1/3} \, ,
\end{eqnarray*}
where
\[ - \frac{u}{2} + \frac{\sqrt{k}}{2\sqrt{6}} \tanh^{-1}(u \sqrt{2/3k}) = \frac{32}{3}t + l \]
and $l$ is another constant. We present some numerical solutions of the system for various 
initial conditions in figure 4.

\subsection{$Nil$ geometry}

When we take $\lambda=-1$, $\mu=\nu=0$, $G$ is isomorphic to the Heisenberg group. 
The flow equations
\begin{equation} \frac{dA}{dt}=-8A^{4},\quad
\frac{dB}{dt}=4BA^{3}, \quad \frac{dC}{dt}=4CA^{3},
\end{equation}
can be easily integrated to yield
\begin{equation}
A=(24t+A_{0}^{-3})^{-1/3},\quad
B=B_{0}\sqrt{A_{0}}A^{-1/2},\quad
C=A^{-1/2}/(B_{0}\sqrt{A_{0}}).
\end{equation}
Therefore, $A \to 0$ whereas $B$ and $C$ diverge. In the
terminology of \cite{IJ}, this degeneracy is called a ``pancake
degeneracy''.

\subsection{$\mathbb{R}^{3}$}
In this case, $\lambda=\mu=\nu=0$, the metric is flat and the flow
equations are trivial.

\subsection{$H^{3}$, $S^{2}\times \mathbb{R}$, and $H^{2}\times
\mathbb{R}$}

These cases do not arise from the construction that uses a $3$
dimensional Lie group as described, however the metrics can be
written explicitly for each case. The metrics for the
three geometries are of the form
$g=\Lambda(t) g_{H^{3}}$, $g=D(t)g_{\mathbb{R}}+E(t)g_{S^{2}}$
and $g=D(t)g_{\mathbb{R}}+E(t)g_{H^{2}}$, where $g_{H^{3}}$, $g_{S^{2}}$, $g_{\mathbb{R}}$
and $g_{H^{2}}$ are the standard metrics on the hyperbolic
3-space, the unit 2-sphere, the real line and the hyperbolic
2-space, respectively. Each of the resulting metrics is already
conformally flat, therefore they are fixed points of the Cotton
flow. Note, however that the Ricci flow degenerates $H^{2}\times \mathbb{R}$ 
and $S^{2}\times \mathbb{R}$ \cite{IJ}. 

\section{Conclusions}

We have introduced a new flow in three dimensions, whose fixed points are conformally flat metrics. We have presented a gradient formulation of this flow by finding an entropy functional. We have also shown that the commutator of this flow with the Yamabe flow, which preserves the conformal class of a metric, is another flow preserving conformal class. We have studied the evolution of the nine homogenous geometries in detail. Four of these, $\mathbb{R}^{3}$, $H^{3}$, $H^{2}\times \mathbb{R}$, $S^{2}\times \mathbb{R}$, are fixed under the flow. Note that the last two of these are degenerated by the Ricci flow. Every homogenous metric in the $SU(2)$ class converges to the round $S^{3}$ metric. For the $Isom(\mathbb{R}^{2})$, $Solv$ and $Nil$ cases, the equations can be solved analytically. Every metric in the $Isom(\mathbb{R}^{2})$ class evolves to a flat metric. The metrics in the $Nil$ class tend to a pancake degeneracy. The metrics in the $SL(2,\mathbb{R})$ and $Solv$ classes develop cigar degeneracies.

There are several points which require further study. As in the Ricci flow case, one can define solitons: 
\begin{equation} 
C_{ij} + \nabla_{i} \xi_{j} + \nabla_{j} \xi_{i} + \lambda g_{ij} = 0. 
\end{equation}
A metric satisfying this equality, which we call a ``Cotton soliton'', can be thought of as evolving under the Cotton flow by just a diffeomorphism and/or a scaling of the underlying manifold. In the case of a
compact manifold, the conservation of total volume implies that $\lambda$ vanishes; however, to allow
for non-compact manifolds, we keep it. We have not been able to find a non-trivial solution of this equation, or of the simpler gradient Cotton soliton equation where $\xi_{j}=\nabla_{j}\varphi$. However,
it may be useful to consider the techniques introduced in \cite{lott,glick} in this regard. 
Another open question is to determine the conditions on the initial metric for which the short time existence problem can be solved, but this seems to be much more difficult than in the second order case. One obviously interesting avenue, for geometric purposes, is the study of singularities formed under the Cotton flow, or under other possible flows obtained by combining the Ricci and Cotton flows. It would also be interesting to find a nonlinear sigma model whose beta function is the Cotton tensor. 

\appendix

\section{Computation of the flow equations}
Here we show the details of the calculation of the Cotton tensor and how (\ref{homogenousflow})
is derived using the differential forms. The connection 1-forms $\omega^{a}\,_{b}$ can be computed using the vanishing of torsion: $de^{a}+\omega^{a}\,_{b}\wedge e^{b}=0$. For the metric (\ref{homogenousmetric}) they specifically take the following form:
\begin{eqnarray}
\omega^{1}\,_{2}=\frac{1}{\sqrt{ABC}}(\lambda A+&\mu B-\nu
C)e^{3}, \;\;
\omega^{1}\,_{3}=\frac{1}{\sqrt{ABC}}(-\lambda A+\mu B-\nu
C)e^{2}, \nonumber\\
&\omega^{2}\,_{3}=\frac{1}{\sqrt{ABC}}(-\lambda A+\mu B+\nu
C)e^{1}.
\end{eqnarray}
Using the formula $R^{a}\,_{b}=d\omega^{a}\,_{b}+\omega^{a}\,_{c}\wedge \omega^{c}\,_{b}$, one  computes the curvature 2-forms as
\begin{eqnarray}
\hspace{-1cm}
R^{1}\,_{2}=\frac{1}{ABC}(\lambda^{2}A^{2}+\mu^{2}B^{2}-3\nu^{2}C^{2}-2\lambda\mu AB+2\lambda\nu AC+2\mu \nu BC)e^{1}\wedge e^{2},\nonumber\\
\hspace{-1cm}
R^{1}\,_{3}=\frac{1}{ABC}(\lambda^{2} A^{2}+\nu^{2}C^{2}-3\mu^{2}B^{2}-2\lambda\nu AC+2\lambda \mu AB+2\mu \nu BC)e^{1}\wedge e^{3},\nonumber\\
\hspace{-1cm}
R^{2}\,_{3}=\frac{1}{ABC}(\mu^{2}B^{2}+\nu^{2}C^{2}-3\lambda^{2}A^{2}-2\mu\nu
BC+2\lambda\mu AB+2\lambda\nu AC)e^{2}\wedge e^{3}.\nonumber \\
\end{eqnarray}
Next, we compute the Ricci 1-forms and the curvature scalar using the formulas $(Ric)_{a}=\iota_{b}R^{b}\,_{a}$ and $\mathcal{R}=\iota_{a}(Ric)^{a}$, respectively, as
\begin{eqnarray}
(Ric)_{1}=\frac{2}{ABC}(\lambda^{2}A^{2}-\mu^{2}B^{2}-\nu^{2}C^{2}+2\mu\nu BC)e^{1},\nonumber\\
(Ric)_{2}=\frac{2}{ABC}(\mu^{2}B^{2}-\nu^{2}C^{2}-\lambda^{2}A^{2}+2\lambda\nu AC)e^{2},\nonumber\\
(Ric)_{3}=\frac{2}{ABC}(\nu^{2}C^{2}-\lambda^{2}A^{2}-\mu^{2}B^{2}+2\lambda\mu
AB)e^{3},\nonumber\\
\mathcal{R}=\frac{2}{ABC}(-\lambda^{2}A^{2}-\mu^{2}B^{2}-\nu^{2}C^{2}+2\lambda\mu
AB+2\lambda \nu AC+2\mu \nu BC).
\end{eqnarray}
The Cotton 2-form is given by $C^{a}=dY^{a}+\omega^{a}\,_{b}\wedge Y^{b}$, where
$Y^{a}=(Ric)^{a}-\frac{1}{4}\mathcal{R}e^{a}$, which leads to
\begin{eqnarray}
\hspace{-1cm}
C^{1}=\frac{4}{(ABC)^{3/2}}[-\lambda^{2}A^{2}(\mu B+\nu C-2\lambda A)-(\mu B+\nu C)(\mu B-\nu C)^{2}]e^{2}\wedge e^{3}, \nonumber\\
\hspace{-1cm}
C^{2}=\frac{4}{(ABC)^{3/2}}[-\mu^{2}B^{2}(\nu C+\lambda A-2\mu B)-(\nu C+\lambda A)(\nu C-\lambda A)^{2}]e^{3}\wedge e^{1}, \nonumber\\
\hspace{-1cm}
C^{3}=\frac{4}{(ABC)^{3/2}}[-\nu^{2}C^{2}(\lambda A+\mu B-2\nu
C)-(\lambda A+\mu B)(\lambda A-\mu B)^{2}]e^{1}\wedge e^{2}. \nonumber \\
\end{eqnarray}
Finally, using $\partial_{t}e^{a}=*C^{a}$, one obtains (\ref{homogenousflow}). 

\section{Coordinate invariant computation of the Cotton tensor}
It is sometimes useful to consider the coordinate free version of the computation above. For this purpose, we follow \cite{Bel} and \cite{CK}. Define the $(0,2)$ tensor $s$ by the
formula:
\begin{equation}
s=Ric-\frac{\mathcal{R}}{4}g. 
\end{equation}
A Cotton 3-form on a generic $D$-dimensional manifold $\mathcal{M}$ can be defined by
\begin{equation}
C(X,Y)(Z)=(\nabla_{X}s)(Y,Z)-(\nabla_{Y}s)(X,Z),
\end{equation}
where $X,Y,Z$ are vector fields on $\mathcal{M}$. Let us now return to the case of 
a homogeneous 3-dimensional manifold. Recall $\{F_{1},F_{2},F_{3}\}$ from section \ref{hofl}. 
First, we compute $s$ using the formulas for $Ric(F_{i},F_{j})$ and $\mathcal{R}$ directly 
from \cite{CK}: 
\begin{eqnarray}
s(F_{1},F_{1})=&\frac{2}{BC}((\lambda^{2}A^{2}-(\mu B-\nu
C)^{2})-\frac{1}{4}(2\mu\nu BC+2\nu\lambda AC+2\lambda \mu AB \nonumber\\
&-\lambda^{2}A^{2}-\mu^{2}B^{2}-\nu^{2}C^{2})),\nonumber\\
s(F_{2},F_{2})=&\frac{2}{AC}((\mu^{2}B^{2}-(\lambda
A-\nu
C)^{2})-\frac{1}{4}(2\mu\nu BC+2\nu\lambda AC+2\lambda \mu AB\nonumber\\
&-\lambda^{2}A^{2}-\mu^{2}B^{2}-\nu^{2}C^{2})), \nonumber\\
s(F_{3},F_{3})=&\frac{2}{AB}((\mu^{2}C^{2}-(\lambda
A-\mu
B)^{2})-\frac{1}{4}(2\mu\nu BC+2\nu\lambda AC+2\lambda \mu AB\nonumber\\
&-\lambda^{2}A^{2}-\mu^{2}B^{2}-\nu^{2}C^{2})) 
\end{eqnarray}
and clearly, $s(F_{i},F_{j})=0$ if $i\neq j$. One has
\begin{equation}
(\nabla_{X}s)(Y,Z)=\partial_{X}(s(Y,Z))-s(\nabla_{X}Y,Z)-s(Y,\nabla_{X}Z).
\end{equation}
When $\{X,Y,Z\}\subset \{F_{1},F_{2},F_{3}\}$, the first
term on the right is $0$ by homogeneity.
$\nabla_{F_{i}}F_{i}=0$ for any $i$, and
\begin{eqnarray}
\nabla_{F_{1}}F_{2}&=\frac{1}{2}\Big([F_{1},F_{2}]-(adF_{1})^{*}F_{2}-(adF_{2})^{*}F_{1}\Big)=\frac{-\lambda A+\mu B+\nu C}{C}F_{3}, \nonumber\\
\nabla_{F_{2}}F_{3}&=\frac{-\mu B+\nu C+\lambda
A}{A}F_{1},\quad
\nabla_{F_{3}}F_{1}=\frac{-\nu C+\lambda A+\nu B}{B}F_{2}, \nonumber \\
\nabla_{F_{2}}F_{1}&=\frac{\mu B-\lambda A-\nu
C}{C}F_{3}, \quad
\nabla_{F_{3}}F_{2}=\frac{\nu C-\mu B-\lambda A}{A}F_{1},\nonumber\\
\nabla_{F_{1}}F_{3}&=\frac{\lambda A-\nu C-\mu B}{B}F_{2}.
\end{eqnarray}
From the definition, $C(X,Y)(Z)$ is antisymmetric in $(X,Y)$, we have
$C(F_{i},F_{i})(F_{j})=0$ for every $i,j$. Since the other antisymmetry is not apparent, we prove: 
\begin{Lem}
For any $i,j$, $C(F_{i},F_{j})(F_{j})=0$.
\end{Lem}
\noindent\textit{Proof:} We may assume $i\neq j$ since the
equality case is taken care of above. Let $k$ be the third index,
different from $i$ and $j$.  Recall that $s$ is diagonal. Now,
\begin{eqnarray}
\hspace{-1cm}
C(F_{i},F_{j})(F_{j})&=(\nabla_{F_{i}}s)(F_{j},F_{j})-(\nabla_{F_{j}}s)(F_{i},F_{j})\nonumber\\
&=-s(\nabla_{F_{i}}F_{j},F_{j})-s(F_{j},\nabla_{F_{i}}F_{j})+s(\nabla_{F_{j}}F_{i},F_{j})
+s(F_{i},\nabla_{F_{j}}F_{j})\nonumber\\
&=s(F_{j},aF_{k})=0,
\end{eqnarray}
where $a$ is some constant. Note that we have again used homogeneity to drop
terms that involve $\partial_{F_{i}}$. This finishes the proof. 

Therefore, the only interesting components
of the tensor are the ones involving all $3$ indices:
\begin{eqnarray*}
\hspace{-1cm}
C(F_{1},F_{2})(F_{3})=-s(\nabla_{F_{1}}F_{2},F_{3})-s(F_{2},\nabla_{F_{1}}F_{3})
+s(\nabla_{F_{2}}F_{1},F_{3})+s(F_{1},\nabla_{F_{2}}F_{3})\\
\hspace{-1cm}
=-s(\frac{-\lambda A+\mu B+\nu
C}{C}F_{3},F_{3})-s(F_{2},\frac{\lambda A-\nu C-\mu
B}{B}F_{2})\\
\hspace{-1cm}
\quad +s(\frac{\mu B-\lambda A-\nu
C}{C}F_{3},F_{3})+s(F_{1},\frac{-\mu B+\nu C+\lambda A}{A}F_{1}) \\
\hspace{-1cm}
=-2\nu s(F_{3},F_{3})-\frac{\lambda A-\nu C-\mu
B}{B}s(F_{2},F_{2})+\frac{-\mu B+\nu C+\lambda
A}{A}s(F_{1},F_{1}),
\end{eqnarray*}
and the other $C(F_{i},F_{j})(F_{k})$ can be computed in a similar
way. Defining $C_{ijk}=C(F_{i},F_{j})(F_{k})$, the flow equations 
(\ref{homogenousflow}) can be written as: 
\begin{equation}
\partial_{t} g_{ij} = \frac{1}{2\sqrt{g}} \, \epsilon^{knm} \, C_{kni} \, g_{jm}.
\end{equation}

In fact, in a general coordinate frame, to relate this formulation to the one 
presented in the main text, we define 
\begin{eqnarray*} 
C_{ijk} & = C(\partial_{i},\partial_{j})(\partial_{k}) \\
& = (\nabla_{i} s)(\partial_{j},\partial_{k})-
(\nabla_{j}s)(\partial_{i},\partial_{k}), \\
& = \partial_{i} (s(\partial_{j},\partial_{k})) - s(\nabla_{i} \partial_{j}, \partial_{k}) 
- s(\partial_{j}, \nabla_{i} \partial_{k}) \\
& \quad - \partial_{j} (s(\partial_{i},\partial_{k})) + s(\nabla_{j} \partial_{i}, \partial_{k}) + s(\partial_{i}, \nabla_{j} \partial_{k}) \, .
\end{eqnarray*}
Using \( s(\nabla_{i} \partial_{j}, \partial_{k}) 
= s(\Gamma^{\ell}\,_{ij} \, \partial_{\ell}, \partial_{k}) 
= \Gamma^{\ell}\,_{ij} \, s(\partial_{\ell}, \partial_{k}), \) this simplifies to
\[ C_{ijk} = \nabla_{i} s_{jk} - \nabla_{j} s_{ik} , \]
where $s_{jk} = s(\partial_{j},\partial_{k})$. In the case of 3 dimensions, one can 
get a symmetric, traceless and covariantly conserved 2-tensor out of this by contracting
it with the completely antisymmetric Levi-Civita tensor:
\[ C_{ij} = \frac{1}{2\sqrt{g}} \, \epsilon^{knm} \, C_{kni} \, g_{jm} \, . \]

\ack
This work is partially supported by the Scientific and Technological Research
Council of Turkey (T{\"U}B\.{I}TAK). B.T. is also partially supported by the
``Young Investigator Grant" of the Turkish Academy of Sciences (T{\"U}BA), and 
both A.U.\"{O}.K. and B.T. are supported by the T{\"U}B\.{I}TAK Kariyer Grant No 104T177.
B.T. would also like thank M. Headrick and A. Waldron for a discussion and 
the U.C. Davis Math. Department where early stages of this work took place. We would
also like to thank a very conscientious referee for useful remarks.


\section*{References}

\begin{thebibliography}{50}
\bibitem{hamilton1} Hamilton R.S.,
  ``Three-manifolds with positive Ricci curvature,''
   J. Diff. Geom. {\bf 17}, 255 (1982).
\bibitem{perelman1} Perelman G.,
  ``The entropy formula for the Ricci flow and its geometric applications,''
  arXiv:math/0211159.
\bibitem{perelman2} Perelman G.,
  ``Ricci flow with surgery on three-manifolds,''
  arXiv:math/0303109.
\bibitem{frie} Friedan D.,
  ``Nonlinear models in two+epsilon dimensions,''
  Phys. Rev. Lett. {\bf 45}, 1057 (1980).
\bibitem{eels} Eells Jr. J. and Sampson J.H.,
  ``Harmonic mappings of Riemannian manifolds,'' Amer. J. Math. {\bf 86}, 109 (1964).
\bibitem{hamilton2} Hamilton R.S.,
  ``The Ricci flow on surfaces,'' in Mathematics and General Relativity (Santa Cruz, CA, 1986) 
  237--262, Contemp. Math. {\bf 71}, Amer. Math. Soc., Providence, RI (1988).
\bibitem{cqgconf} Fischer A.E.,
  ``An introduction to conformal Ricci flow,''
  Class. Quantum Grav. {\bf 21}, S171 (2004).
\bibitem{gegenberg} Gegenberg J. and Kunstatter G.,
  ``Using 3D string-inspired gravity to understand the Thurston conjecture,''
  Class. Quantum Grav. {\bf 21}, 1197 (2004).
\bibitem{headrick} Headrick M. and Wiseman T.,
  ``Ricci flow and black holes,''
  Class. Quantum Grav. {\bf 23}, 6683 (2006),
  [arXiv:hep-th/0606086].
\bibitem{samuel} Samuel J. and Chowdhury S.R.,
  ``Energy, entropy and the Ricci flow,''
  Class. Quantum Grav. {\bf 25}, 035012 (2008),
  [arXiv:0711.0430 [gr-qc]].
\bibitem{woolgar} Woolgar E.,
  ``Some applications of Ricci flow in physics,''
  arXiv:0708.2144 [hep-th].
\bibitem{Cot} Cotton E.,
  ``Sur les vari\'et\'es \`a trois dimensions,''
  Ann. Fac. d. Sc. Toulouse (II) {\bf 1}, 385 (1899).
\bibitem{Eis} Eisenhart L.P., 
  {\it Riemannian Geometry}, 
  (Princeton Univ. Press, Princeton, NJ), (1997) (8th printing) pp 91.
\bibitem{adm} Arnowitt R., Deser S. and Misner C.W., 
  Phys. Rev. {\bf 117}, 1595 (1960); in {\it Gravitation: An 
  Introduction to Current Research}, ed. by L. Witten (Wiley, NY, 1962); 
  ``The dynamics of general relativity,'' [arXiv:gr-qc/0405109].
\bibitem{york} York J.W.,
  ``Role of conformal three geometry in the dynamics of gravitation,''
  Phys. Rev. Lett. {\bf 28}, 1082 (1972).
\bibitem{garcia}
  A. Garcia, F.W. Hehl, C. Heinicke and A. Macias,
  ``The Cotton tensor in Riemannian spacetimes,''
  Class. Quantum Grav. {\bf 21}, 1099 (2004),
  [arXiv:gr-qc/0309008].
\bibitem{DJT} Deser S., Jackiw R. and Templeton S.,
  ``Topologically massive gauge theories,''
  Annals Phys. {\bf 140}, 372 (1982).
\bibitem{DT} DeTurck D.M.,
  ``Deforming metrics in the direction of their Ricci tensors,''
  J. Diff. Geom. {\bf 18}, 157 (1983).
\bibitem{Thu} Thurston W.,
  ``Three dimensional manifolds, Kleinian groups, and hyperbolic geometry,''
  Bull. Amer. Math. Soc. (N.S.) {\bf 6}, 357 (1982).
\bibitem{IJ} Isenberg J. and Jackson M., ``Ricci flow on locally homogeneous geometries on closed manifolds,'' J. Diff. Geom. {\bf 35}, 723 (1992).
\bibitem{KM} Knopf D. and McLeod K., ``Quasi-convergence of model geometries under the Ricci flow,'' Anal. Geom. {\bf 9}, 879 (2001).  
\bibitem{CK} Chow B. and Knopf D., {\em The Ricci flow: An introduction}, Math. Surveys 
and Monographs {\bf 110}, Amer. Math. Soc. (2004).
\bibitem{caoni} Cao X., Ni Y. and Saloff-Coste L., ``Cross curvature flow on locally homogenous three-manifolds (I),'' [arXiv:0708.1922 [math.DG]]
\bibitem{Mil} Milnor J., ``Curvatures on left invariant metrics on Lie groups,'' Adv. in Math. {\bf 21}, 293 (1976).
\bibitem{lott} Lott, J., ``On the long-time behavior of type-III Ricci flow solutions,'' Math. Ann.  
{\bf 339}, 627 (2007).
\bibitem{glick} Glickenstein D., ``Riemannian groupoids and solitons for three-dimensional 
homogeneous Ricci and cross-curvature flows,'' to appear in Int. Math. Res. Not. (2008).
\bibitem{Bel} Belgun F.A., ``Null-geodesics in complex conformal manifolds and the LeBrun correspondence,'' J. Reine Agnew. Math. {\bf 536}, 43 (2001).
\end{thebibliography}
\end{document}